\documentclass{svproc}
\usepackage{bm} 
\usepackage{amsmath}
\usepackage{amsfonts}
\usepackage{amssymb}
\usepackage{subcaption}
\usepackage{booktabs}
\usepackage{graphicx}

\usepackage{array}

\usepackage[numbers,sort&compress,sectionbib]{natbib}

\newcommand{\x}{\mathbf{x}}
\newcommand{\rr}{\mathbf{r}}

\newcommand{\kcal}{kcal~mol$^{-1}$}

\usepackage{url}

\graphicspath{{./figures/}}

\begin{document}
\mainmatter              
\title{Quantum-chemical insights from interpretable atomistic neural networks}
\titlerunning{Quantum chemical insights}  
%
\author{Kristof T. Sch\"utt\inst{1} \and Michael Gastegger\inst{1} \and Alexandre Tkatchenko\footnote[1]{Corresponding authors: \\ alexandre.tkatchenko@uni.lu \\ klaus-robert.mueller@tu-berlin.de}\inst{2} \and Klaus-Robert M\"uller$^*$\inst{1,3,4}}
\authorrunning{K. T. Sch\"utt et al.} 
%
\tocauthor{Kristof T. Sch\"utt, Michael Gastegger, Alexandre Tkatchenko, and Klaus-Robert M\"uller}

\institute{Technische Universit\"at Berlin, 10587 Berlin, Germany, Machine Learning Group
\and
Physics and Materials Science Research Unit, University of Luxembourg, L-1511 Luxembourg, Luxembourg
\and
Max-Planck-Institut f\"ur Informatik, Saarbr\"ucken, Germany
\and
Department of Brain and Cognitive Engineering, Korea University, Anam-dong, Seongbuk-gu, Seoul 136-713, South Korea}

\maketitle              

\begin{abstract}
With the rise of deep neural networks for quantum chemistry applications, there is a pressing need for architectures that, beyond delivering accurate predictions of chemical properties, are readily interpretable by researchers.
Here, we describe interpretation techniques for atomistic neural networks on the example of Behler--Parrinello networks as well as the end-to-end model SchNet.
Both models obtain predictions of chemical properties by aggregating atom-wise contributions.
These latent variables can serve as local explanations of a prediction and are obtained during training without additional cost.
Due to their correspondence to well-known chemical concepts such as atomic energies and partial charges, these atom-wise explanations enable insights not only about the model but more importantly about the underlying quantum-chemical regularities.
We generalize from atomistic explanations to 3d space, thus obtaining spatially resolved visualizations which further improve interpretability.
Finally, we analyze learned embeddings of chemical elements that exhibit a partial ordering that resembles the order of the periodic table.
As the examined neural networks show excellent agreement with chemical knowledge, the presented techniques open up new venues for data-driven research in chemistry, physics and materials science.
\end{abstract}

\section{Introduction}
The discovery of novel molecules and materials is crucial for research in a wide variety of applications ranging from food processing and drug design~\cite{shoichet2004virtual,bajorath2002integration} to more efficient batteries~\cite{Kang2006lithiumbattery,chen2012lithiumbattery,hautier2013lithiumbattery} and solar cells~\cite{Olivares2011organicphotovoltaics}.
While quantum-chemical calculations~\cite{KS, DFT} deliver the means to predict such properties for given atomistic systems, their computational cost as well as the vastness of chemical compound space prevents an exhaustive exploration~\cite{von2013first}.
In recent years, there has been a growing interest in applying machine learning techniques to model quantum-chemical systems~\cite{rupp2012fast,montavon2013machine,Hansen-JCPL,hansen2013assessment,schutt2014represent,bartok2010gaussian,behler2007generalized,chmiela2017machine,faber2017fast,eickenberg2017solid,brockherde2017bypassing}.
While research has focused primarily on predicting chemical properties by applying non-linear regression methods such as Gaussian processes or neural networks to manually crafted features\cite{bartok2015gaussian,behler2014representing}, there have also been successful approaches to learn molecular representations end-to-end.
These include neural circular fingerprints that use chemical graphs as inputs~\cite{duvenaud2015convolutional,kearnes2016molecular}, mixed approaches that use both graph information as well as atomic positions~\cite{gilmer2017neural} and architectures that learn purely from first-principles information such as deep tensor neural networks (DTNNs)~\cite{schutt2017quantum}, which represent atomistic systems by modeling subsequent pair-wise interactions of atomic environments with factorized tensor layers.
Other architectures fitting into the DTNN framework include SchNet~\cite{schutt2018jcpschnet}, where the interactions are modeled using continuous-filter convolutions~\cite{schutt2017schnet} as well as more recent variations of this theme such as HIP-NN~\cite{lubbers2018hierarchical} or crystal graph convolutional networks~\cite{xie2018crystal}.

As these neural network architectures become increasingly complex, it is crucial that quantum-chemistry researchers are able to acquire an intuition how these models function and how trustworthy predictions are.
Beyond a high prediction accuracy, this requires neural networks to demonstrate that they have learned fundamental quantum-chemical principles.
Several techniques have been developed that generate explanations for classifier decisions of neural networks~\cite{baehrens2010explain,Bach2015,Montavon2017,Simonyan2013,Zeiler2014,kindermans2018learning,zintgraf2017visualizing}.
Since quantum-chemical properties are often continuous, such as the prediction of molecular energies with a neural network potential, regression problems are more common in this field than classification.
This changes how explanations have to be interpreted.
Given a neural network potential, saliency maps based on input gradients~\cite{baehrens2010explain, Simonyan2013} correspond to the force that acts on atoms.
While this might indeed be a reason for high energies, e.g. if two atoms are very close, the gradient is too local to explain the energy level sufficiently.
This is especially the case for stable (equilibrium) molecules, which are located in a local energy minimum such that all forces are zero.
Therefore, input gradients would indicate that the atom positions are not important, which is clearly wrong.
Other explanation methods assign importance or relevance scores to input features through obtaining reverse mappings based on the network parameters~\cite{Bach2015,Montavon2017,Zeiler2014}, sampling~\cite{zintgraf2017visualizing} or training for signal reconstruction~\cite{kindermans2018learning}.
Even though some of those alleviate the problem of pure input gradients since their explanations are less local~\cite{samek2017evaluating}, there is another fundamental issue in this application:
While pixel-wise relevance scores of images allow for an intuitive inspection of the obtained explanation, the influence of the positions and types of individual atoms is not readily interpretable in the quantum chemical picture.
Here, we aim for an explanation in the full 3-d space, i.e. beyond positions of nuclear charges.

In the following, we will introduce two neural network potentials, namely (1) Behler-Parrinello networks (BP)~\cite{behler2007generalized,behler2011atom,gastegger2018wacsf} that make use of manually engineered features and (2) SchNet~\cite{schutt2017schnet,schutt2018jcpschnet}, which learns atomistic representations directly from atom types and positions.
For both architectures, we will demonstrate interpretation strategies that allow for spatially and chemically resolved insights into the inner workings of the neural network as well as the underlying data.
Furthermore, we will show that both kinds of architectures -- and deep end-to-end models in particular -- not only are highly accurate, but recover fundamental chemical knowledge.

\section{Atomistic Neural Network Potentials}

Due to the spatial structure of atomistic systems and the nature of quantum mechanical laws giving rise to various invariances and scaling behaviors of chemical properties, special adaptations to conventional neural network architectures are necessary in order to model chemical systems efficiently.
The first major issue arises from the overall diversity exhibited by molecules.
They can vary greatly with respect to the overall number of atoms as well as the combination of chemical elements present, thus rendering purely static architectures ill-suited for obtaining a general description.
In addition, molecular properties do not change if atoms of the same element are exchanged and the corresponding invariances with respect to atom types needs to be accounted for by the model.

Second, the properties of molecules originate from interactions between nuclei and electrons.
These can be roughly represented by interatomic potentials which are functions in 3d space depending on the types and positions of the atoms.
However, atom coordinates -- and subsequently all associated molecular properties -- can change in a continuous manner.
Hence, all grid based methods (e.g. conventional convolutional neural networks) are generally infeasible, as they fail to resolve these incremental changes. 
Moreover, chemical properties are invariant with respect to translations and rotations in Cartesian space, imposing additional constraints on machine learning models for molecules and materials.

In order to overcome the first of the above issues, so-called atomistic neural network architectures are introduced. 
Similar to neural networks for graphs, the atomistic system is decomposed into local environments.
Specifically, a set of feature vectors is defined for every atom based on which latent atom-wise contributions to a property of interest are predicted. 
These are used to reconstruct the target property via physically motivated aggregation layers that guarantee permutational invariance of the atoms.

\begin{figure}[tb]
	\centering
	\includegraphics[width=0.7\textwidth]{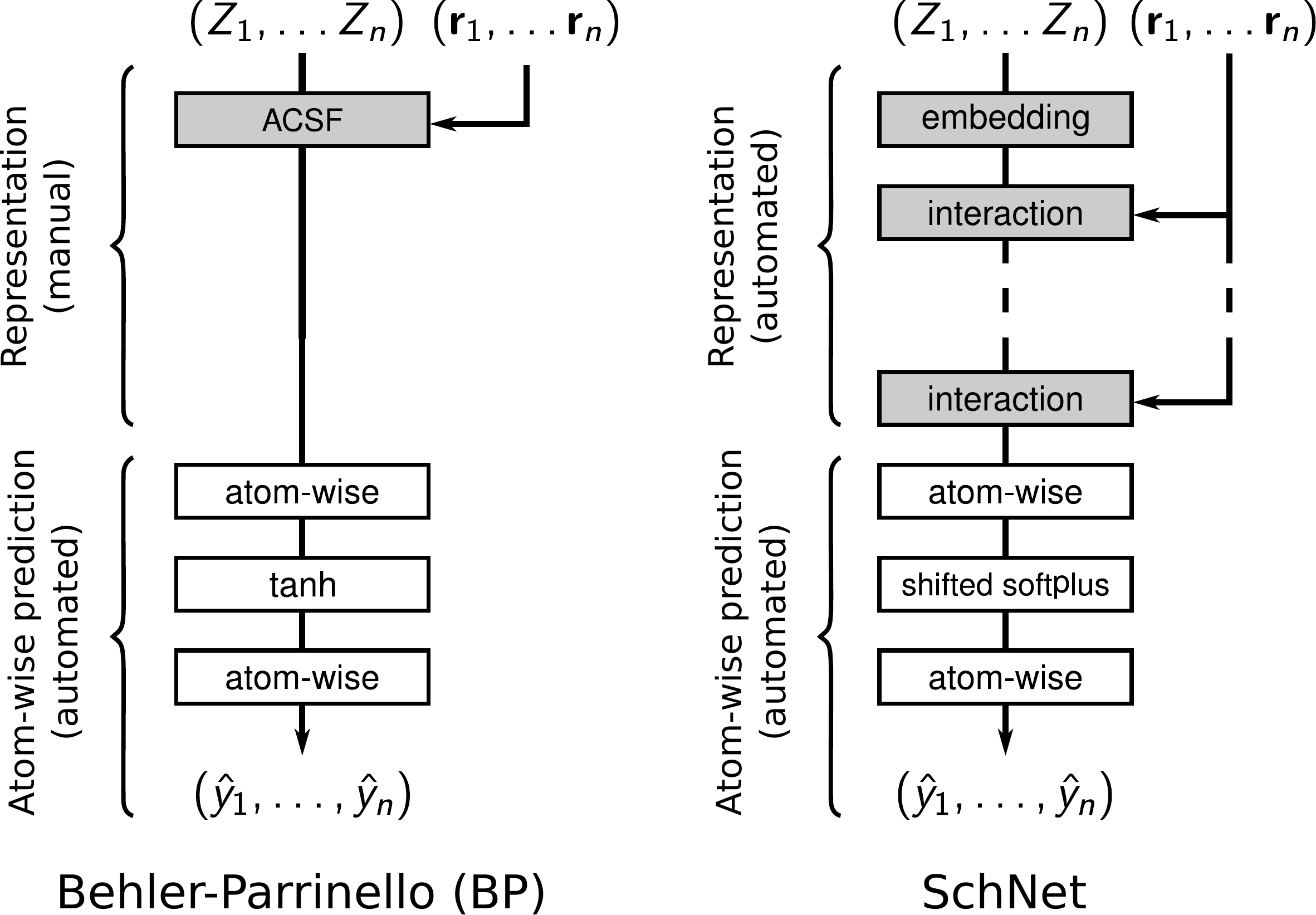}
	\caption{Illustration of the two examined neural network architectures: Behler-Parrinello network with atom-centered symmetry functions (ACSFs, left) and the end-to-end architecture SchNet (right).\label{fig:bp_v_schnet}}
\end{figure}
Depending on the strategy used to obtain atom-wise features, two categories of atomistic neural network models can be distinguished (see Figure~\ref{fig:bp_v_schnet}). 
The first type employs handcrafted features, which are engineered before training. 
A popular choice in this category are Behler--Parrinello (BP) networks using atom-centered symmetry functions~\cite{behler2007generalized,behler2011atom}.
In the second category, all spatial invariances are encoded instead directly into the structure of an atomistic neural network such that atom-wise representations can be obtained during training in an end-to-end fashion.
This includes neural networks implementing the DTNN framework, where atomistic representations are constructed through interaction layers such as the continuous-filter convolutional neural network SchNet~\cite{schutt2017schnet,schutt2018jcpschnet}.

In the following section, BP and SchNet architectures will be discussed in greater detail.
Finally, a short overview will be given on how various chemical properties are obtained in an atomistic machine learning framework.

\subsection{Behler--Parrinello Potentials}

BP neural network potentials apply fully-connected neural networks atom-wise to so-called atom-centered symmetry functions (ACSFs)~\cite{behler2011atom}.
These ACSFs describe the arrangement and chemical identities of the neighbors surrounding a central atom via sets of specialized distribution functions.
Typically, multiple, different types of ACSFs are used to capture radial and angular information.

Radial distribution functions take the form
\begin{equation}
G^\mathrm{rad}_i = \sum_{j\neq i}^N e^{-\eta(r_{ij}-r_0)^2} f_\mathrm{cut}(r_{ij}),
\end{equation}
where $N$ is the number of atoms in the molecule and $r_{ij}$ the distance between the central atom $i$ and its neighbor $j$. 
The parameters $\eta$ and $r_0$ control the width and position of the Gaussian.
The cutoff function $f_\mathrm{cut}$ ensures that the contribution of every neighbor to the ACSF decays to zero if it is located too far away from the central atom.
As radial functions offer only a limited spatial resolution, they are used in combination with angular ACSFs. 

In order to account for different chemical species in an atoms environment, ACSFs are typically defined for pairs (radial) and triples (angles) of chemical elements. 
In addition, a set of radial and angular ACSFs differing in their respective hyper-parameters is used for every resulting combination in order to provide a sufficiently resolved description of chemical environments.
Thus, the number of features and hyper-parameters grows quickly with the number of  chemical elements present in the data set.
However, strategies have been proposed to overcome some of these problems, such as introducing an element-dependent weighting of ACSFs in order to avoid the combinatorial explosion of features~\cite{gastegger2018wacsf}.

Due to the above definition, the hyper-parameters of all individual functions need to be determined in a tedious trial and error procedure based on the molecules under investigation. 
However, ACSFs engineered based on the domain knowledge of a skilled practitioner can be highly efficient in terms of required reference calculations for training~\cite{gastegger2018wacsf}.


\subsection{SchNet}

In contrast to the previously described architecture, SchNet is able to learn an efficient representation of chemical environments directly from atom types and positions with minimal hyper-parameter tuning.
The overall structure of SchNet follows the DTNN framework~\cite{schutt2017quantum} consisting of three steps:
\begin{enumerate}
	\item Initialize atom features $\mathbf{x}_i$ with embeddings of chemical element $Z_i$:
	\[
		\mathbf{x}_i^{(0)} = \mathbf{A}_{Z_i}
	\]
	\item Infuse spatial information of the chemical environment adding pair-wise interaction corrections $\mathbf{v}^{(t)}$ multiple times:
	\[
	\mathbf{x}^{(t+1)}_i = \mathbf{x}^{(t)}_i + \sum\limits_{j \neq i} \mathbf{v}^{(t)} \left(\mathbf{x}^{(t)}_j, r_{ij} \right)
	\]
	\item Obtain property of interest from final atom-wise representations $\mathbf{x}^{(T)}_i$ using physically motivated aggregation (see Sec.~\ref{sec:prop}).
\end{enumerate}

The crucial difference between various implementations of the DTNN framework is how the interaction corrections $\mathbf{v}^{(t)}$ are modeled.
In case of SchNet, we apply a continuous-filter convolution~\cite{schutt2017schnet} over the atomistic system with a smooth convolution filter generated by a fully-connected neural network depending on the pair-wise distances $r_{ij}$:
\[
(\x * W)(\rr_i) = \sum\limits_{j=1}^{N_\text{atom}} \mathbf{x}_j^{(t)} \circ \underbrace{W^{(t)}(r_{ij})}_{\text{filter-generating network}}
\]
Defining the interaction correction $\mathbf{v}^{(t)}$ using such a convolution on pair-wise distances results in radial filters, i.e. rotational and translational invariances are guaranteed.
Due to the repeated interaction corrections, spatial information is propagated across multiple atoms.
Thus, many-body interactions can be inferred without having to explicitly include angular or higher-order information~\cite{behler2007generalized,pronobis2018many,huo2017unified}.


\subsection{Chemical Properties}\label{sec:prop}

In atomistic models, a chemical property is expressed via latent atomistic contributions. Based on these contributions, the original property is then reconstructed via a physically motivated aggregation layer.
The exact functional form strongly depends on the property.

A common target of atomistic machine learning approaches is the atomization energy $E$.
It can be seen as a measure of how stable different molecules and their configurations are compared to each other and allows to make predictions about the reactivity of chemical species.
In an atomistic framework, the aggregation for the potential energy of a molecule takes the form
\begin{equation}
E = \sum^N_i \hat{E}_i, \label{eq:atpot}
\end{equation}
where $\hat{E}_i$ are latent atomic contributions to the energy.
In case of BP and SchNet, they are obtained from atom-wise prediction layers that take the respective atom-wise representations as input.
Due to the summation in Eq.~\ref{eq:atpot}, atomistic models implicitly account for permutation invariance and can be applied to molecules of arbitrary size and composition.

Another chemical property of interest is the dipole moment $\bm{\mu}$ or its magnitude $\mu$~\cite{gastegger2017machine,sifain2018discovering}. 
Those properties are a measure for the separation of regions of positive and negative charge in a molecule and, for instance, important in infrared spectroscopy.
The dipole moment vector $\bm{\mu}$ can be written as
\begin{equation}
\bm{\mu} = \sum^N_i \hat{q}_i \mathbf{r}_i. \label{eq:atmu}
\end{equation}
where $\hat{q}_i$ are latent partial charges predicted from atom-wise representations.
The positions $\mathbf{r}_i$ of atom $i$ are given relative to a reference point, typically the molecules center of mass.
Based on expression~\ref{eq:atmu}, the magnitude of the dipole moment $\mu$ simply is
\begin{equation}
\mu = \left\|\bm{\mu} \right\|_2 = \left\| \sum^N_i \hat{q}_i \mathbf{r}_i \right\|_2.\label{eq:totmu}
\end{equation}

An important feature of atomistic architectures is that the latent properties are not learned directly, but inferred by the neural network. 
Only the molecular energies and dipole moments are quantum-mechanical observables and can hence be computed based purely on first principles. 
Although atomic energies and partial charge distributions can not be derived in a unique manner, they nevertheless constitute important tools to characterize and interpret the properties and behavior of atomistic systems.
In this sense, atomistic models represent a new class of purely data driven partitioning schemes for chemical properties.

\section{Interpretability}
\begin{table}[t]
	\caption{Mean absolute errors and root mean squared errors of analyzed models trained on 100k molecules from the QM9 benchmark dataset.\label{tab:performance}}
	\centering
	\begin{tabular}{llrrrr} 
		\toprule
		& & \multicolumn{2}{c}{\textbf{Behler-Parrinello}} & \multicolumn{2}{c}{\textbf{SchNet}} \\
		\textit{Property} & \textit{Unit} & \makebox[1.7cm][r]{\textit{MAE}} & \makebox[1.7cm][r]{\textit{RMSE}} & \makebox[1.7cm][r]{\textit{MAE}} & \makebox[1.7cm][r]{\textit{RMSE}} \\ \midrule
		Atomization energy & \kcal & 0.77 & 1.32 & {0.35} & {0.94} \\
		Dipole moment & Debye & 0.073 &  0.118 & 0.025 & 0.050 \\ \bottomrule
	\end{tabular}
\end{table}

As stated in the introduction, conventional interpretation techniques work well for neural networks on images or text, however can not sufficiently explain predictions of continuous chemical properties that depend on interatomic potential spanning the whole 3d space.
Instead, we investigate approaches particularly tailored to these kind of problems, exploiting several features of atomistic models in the process.
E.g., analyzing latent contributions of chemical environments to a property of interest opens up new venues for interpreting atomistic neural networks from a machine learning perspective~\cite{schutt2017quantum}.
Moreover, many of these explanation schemes are directly related to physical and chemical properties of the molecules under study, allowing to extract chemical insights from the model.

In the following, we will demonstrate three interpretable aspects of atomistic models, namely (1) atom-wise latent contributions, (2) probing representations in 3-d space and (3) embeddings of chemical elements.
For all of our analyses, we will employ BP and SchNet models trained on 100k reference calculations at the B3LYP level of theory~\cite{lee1988development,becke1988density} from the popular QM9 molecule benchmark~\cite{ramakrishnan2014quantum}.
The dataset consists of all possible molecules with up to nine heavy atoms from the \{C, O, N, F\} set of chemical elements and are chemically saturated with hydrogen~\cite{blum2009gdb13,ruddigkeit2012enumeration}.
Table~\ref{tab:performance} shows the performance of the trained models.
SchNet achieves consistently lower errors since it is able to adapt its representation to the data at hand, while BP employs a fixed feature representation.
This is especially advantageous in the chemical compound space setting with a large and diverse set of training molecules.
On this ground, we will analyze how both models obtain predictions of chemical properties as well as whether the obtained latent variable agree with chemical intuition and can be employed to extract further insight.

%

\subsection{Atom-wise Partitioning of Chemical Properties}\label{sec:distributions}
A major feature of atomistic architectures is the access to atom-wise latent variables, providing a framework for atom-wise explanation out-of-the-box.
This atom-wise saliency can be seen as the logical extension of the pixel-wise explanations used for images to the domain of molecules.
Unlike relevance propagation approaches~\cite{Bach2015,montavon2017methods}, interpretable features are an implicit part of the model architecture and obtained during training without additional cost, similar to approaches for weakly-supervised object detection~\cite{pinheiro2015image}.
The final prediction is aggregated via physically motivated aggregation layers from the latent variable which thereby get assigned inherent physical interpretations.
Since the use of these aggregation layers is not restricted to a particular class of atomistic architectures, valuable information can be gained for any type of model -- independent on whether models use hand-crafted features such as BPs or learn representations end-to-end such as SchNet.
This makes it possible to compare different models at new levels of abstraction, gaining insights into their inner workings and fundamental differences.

\begin{figure}[htb]
	\centering
	\includegraphics[width=\textwidth]{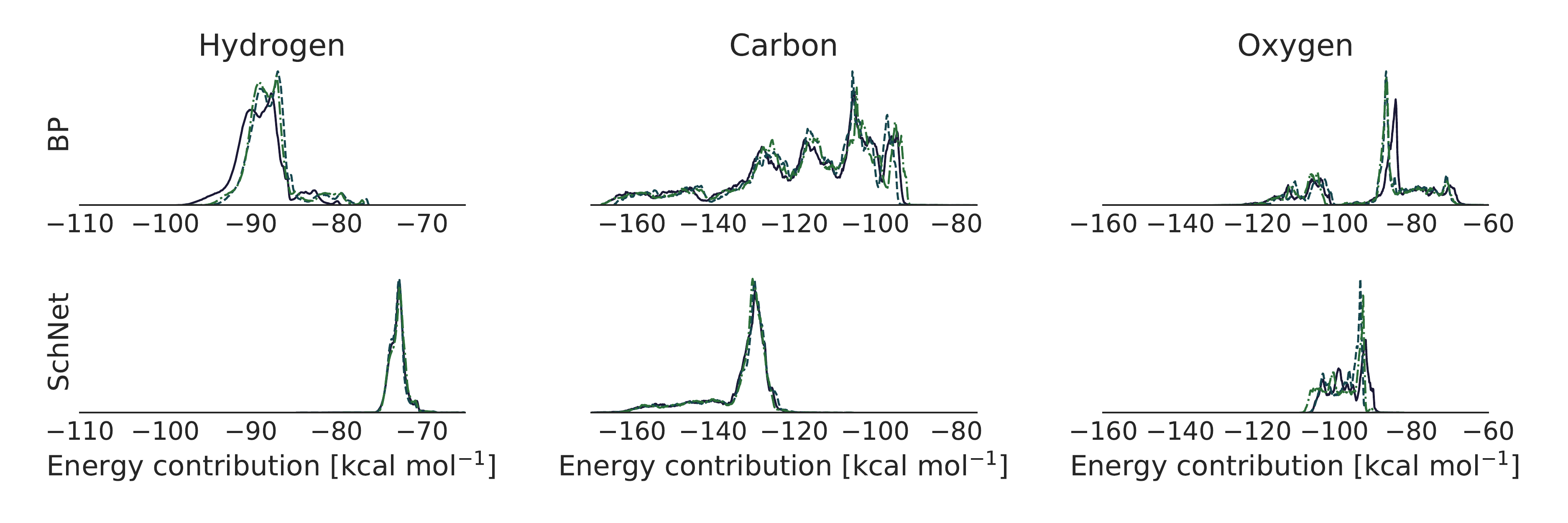}
	\caption{Distribution of energy contributions for atoms of types H, C, N, O from QM9 molecules predicted by Behler-Parrinello and SchNet models. The models were trained on 100k examples. Each color corresponds to a model trained on a different subset.\label{fig:energycontribution}}
\end{figure}

When the property of interest is the atomization energy of an atomistic system, atomic energy contributions are obtained as latent properties.
Figure~\ref{fig:energycontribution} depicts the distributions of these energies obtained for the BP and SchNet models and different folds of the QM9 database.
While the energy contributions within a model are well conserved in general, we find that this effect is significantly more pronounced for the SchNet architecture.
Beyond that, it is possible to discern effects due to the frequency of atom types in the reference data.
Less frequent elements such as oxygen show greater variation compared to the abundant hydrogen and carbon atoms.

As shown in Figure~\ref{fig:energycontribution}, both atomistic models arrive at qualitatively different partitionings of the atomic energies.
The differences observed between the latent variables allow for insight about how energy predictions are obtained.
Generally, energy distributions of the BP architecture are wider than their SchNet counterparts and show more distinct features.
The main reason for this behavior is the way, how both architectures represent molecular structure.
In BP networks, ACSFs are engineered before training to provide a sufficient resolution of different chemical environments.
During the learning process, the atomistic energy contributions are adapted based on these predetermined features, which introduce a certain bias.
Hence, patterns already present in the descriptors are more or less retained in the latent properties.
This is particularly prominent in the case of carbon, where the different peaks of the distribution simply correspond to the various local environments present in QM9.
SchNet on the other hand learns appropriate representations in an end-to-end manner exclusively from the reference data.
The narrow shape observed for the SchNet energy distributions indicates that this type of model arrives at a simple solution of the learning problem by keeping the deviation of the interaction energies within atom types to a minimum.

\begin{figure}
	\centering
	\newcolumntype{S}{>{\centering\arraybackslash} m{.01\textwidth} }
	\newcolumntype{T}{>{\centering\arraybackslash} m{.99\textwidth} }
	\begin{tabular}{ST}
		\rotatebox{90}{SchNet} & \includegraphics[width=\textwidth]{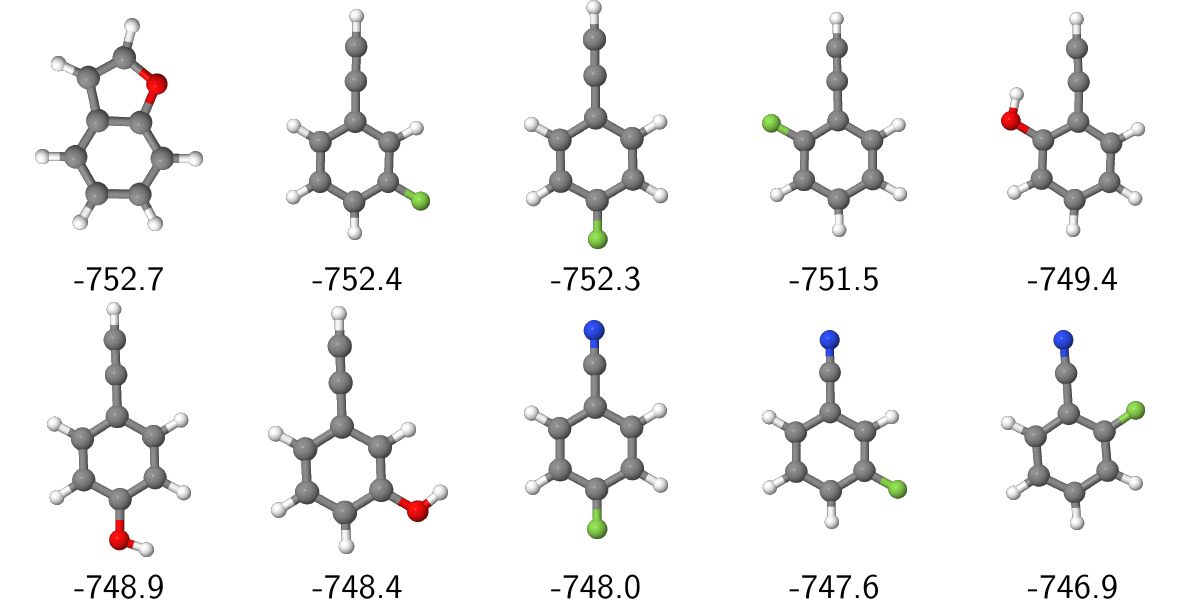} \\
		& \\
		\rotatebox{90}{BP} & \includegraphics[width=\textwidth]{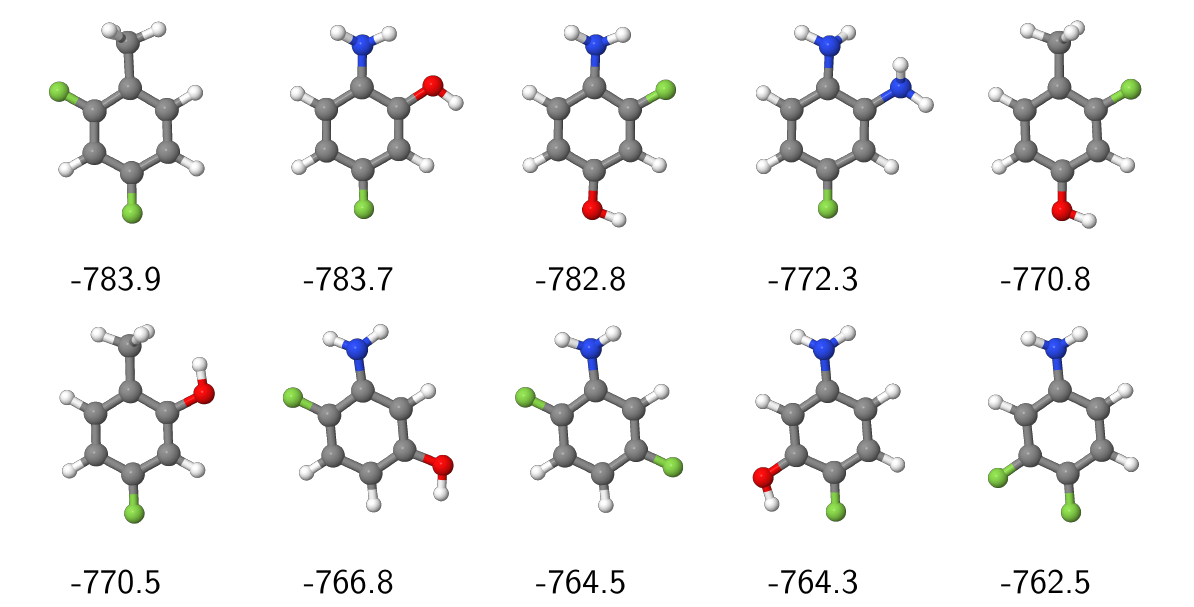} \\
	\end{tabular}
	
	\caption{Energy ranking of 6-membered carbon rings in the QM9 dataset obtained from atom-wise energy contributions as predicted by SchNet (top) and a Behler--Parrinello model (bottom). For each architecture, we show the ten most stable 6-membered carbon rings according to this metric. The atom types are colored as follows: hydrogen--white, carbon--gray, nitrogen--blue, fluorine--green.\label{fig:rings}}
\end{figure}

These atomic energies can also serve as a basis for constructing novel measures of more abstract chemical concepts.
An example for such an application is the use of atomic energies as a stability ranking for aromatic rings with different substitution patterns.
The ten most stable rings in the QM9 database determined in this way are shown in Figure~\ref{fig:rings}.  
The SchNet stability ranking appears to capture central aspects of the chemistry of the investigated systems.
For example, the most stable ring is found to be adjacent to a five membered ring involving oxygen.
Since the carbon atoms in the smaller ring are connected via a double bond, the $\pi$ system of the aromatic ring is extended, leading to the high stability.
This phenomenon is also referred to as the mesomeric effect in organic chemistry~\cite{vollhardt2014organic}.
The same reasoning holds true for alkyne substituents, which are found in six out of the ten structures.
Another common motif is the presence of a fluorine atom.
Due to its high electronegativity, fluorine forms very strong bonds with carbon, thus contributing greatly to the overall stability of the system.
In case of the BP ranking, similar patterns are found for fluorine.
Otherwise, the BP model shows preference for groups donating electron density to the central ring, such as hydroxy (OH) and amine (NH$_2$) groups. 
This trend is referred to as the inductive effect in organic chemistry and is known to increase ring stability similar to the mesomeric effect observed above~\cite{vollhardt2014organic}.
Finally, we find that the BP based model attributes more energy to the ring carbons than SchNet, providing further evidence that SchNet strives to learn a partitioning that minimizes the deviation of the interaction energies within atom types.
This interplay between explaining model predictions via chemical reasoning and obtaining new insights into investigated systems themselves constitutes one of the most tantalizing aspects of applying these methods to physically or chemically motivated problems.

\begin{figure}[tb]
	\centering
	\includegraphics[width=\textwidth]{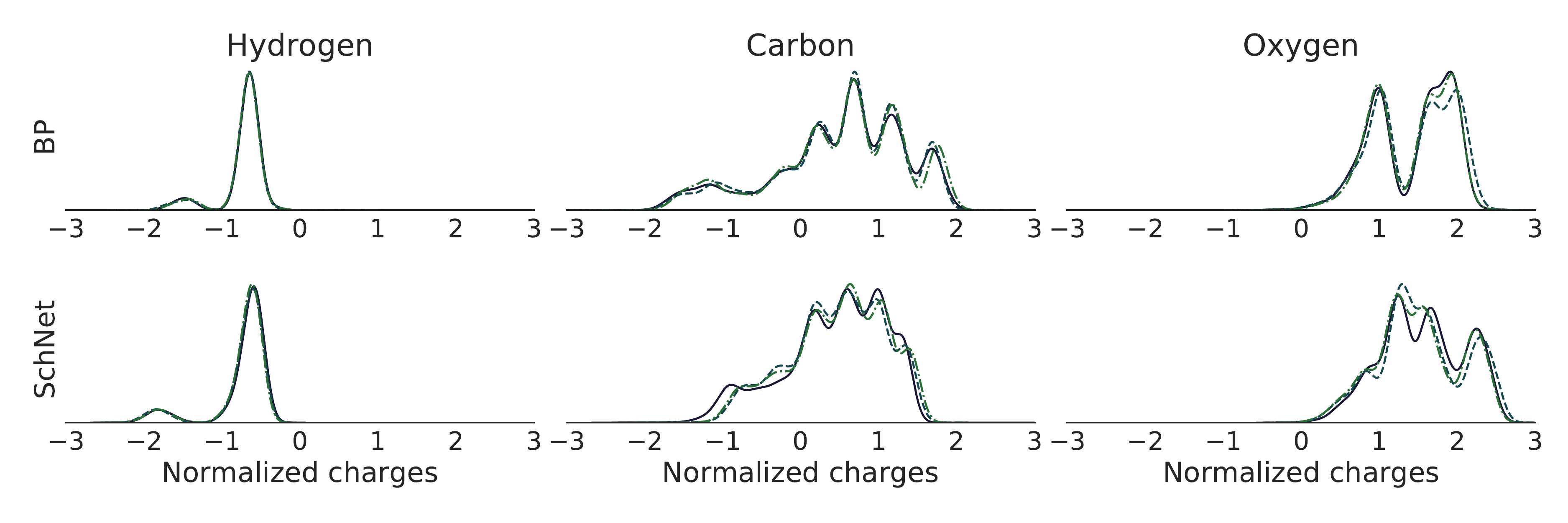}
	\caption{Distributions of latent charges from Behler-Parrinello and SchNet dipole models.\label{fig:chargedistribution}}
\end{figure}

Using the molecular dipole moment as the target property, the atomistic networks yield latent atomic partial charges instead of energies (see Equation~\ref{eq:totmu}).
In direct analogy to the atomic energies, the resulting atom-wise explanations can be used to gain insights not only on a model level, but also on a physical level.
Pertaining to model level insights, qualitative differences between the energy and dipole models, as well as between BP and SchNet architectures, can be elucidated based on the distribution of partial charges obtained for all molecules in QM9 (Fig.~\ref{fig:chargedistribution}).
Comparing the distributions obtained for the same model trained on different subsets of the data, we find that in general the distributions of partial charges are more conserved than those obtained for the atomic energies (Fig.~\ref{fig:energycontribution}). 
The reason for this behavior is the additional structural information present in the dipole aggregation operation (Equation~\ref{eq:totmu}).
The dependence on the atom positions $\mathbf{r}_i$ and hence on the molecular shape introduces additional prior knowledge, thus leading to a more unique partitioning (up to a constant scaling factor).
Further support for this conclusion is offered by the observation that the distribution of charges obtained with BP networks and SchNets shows a much closer agreement than for the atomic energies.
This effect is especially pronounced for the hydrogen and carbon partial charge distributions, which exhibit very similar features.
Analyzing these features for the carbon atom, one also notices parallels between the energy and charge distribution obtained for the BP type model, whereas the SchNet counterparts show little to no similarity.
As stated above, the reason for this phenomenon is the static nature of descriptors employed in BP models, which stay the same irrespective of the target property. SchNet on the other hand is able to infer different, more optimal representations of the molecular structure depending on the modeling task.

In the case of dipole moments and partial charges, interpretation on the physical level takes on particularly interesting characteristics.
The ability to obtain partial charges based exclusively on the dipole moment is remarkable, as it offers insights into the internal structure of a molecule -- in this case the charge distribution -- based on a single global property.
These partial charges can in turn be used to rationalize e.g. chemical reaction mechanisms, molecular reactivity or the aggregation behavior of molecules.
In the next section, we will explore how to visualize such spatially resolved insights.


\subsection{Visualization of Spatially Resolved Insights}

Having inspected atom-wise latent contributions, we will now introduce a feature of the DTNN framework that allows us to extend such atom-wise explanations to interpretable visualizations in 3-d space.
Since energies are obtained atom-wise through a series of pair-wise interaction corrections, it is possible to obtain an energy contribution for every point in space.
To this end, we introduce a test charge $p$ to the atomistic system which we will use to probe the space surrounding the atoms.
This enables us to examine the representation regarding spatial changes and interactions.
In particular, we obtain a more intuitive visualization of the interactions within the molecule, as they haven been learned by the neural network.

Since we only can represent atoms in SchNet, the test charge is bound to be an atom in our model.
This brings the problem that the molecule would be drastically influenced by adding another atom and, moreover, that the resulting molecule is bound to leave the training manifold if we trained the neural network only on equilibrium configuration or single molecular dynamics trajectories with a fixed number of atoms.
We solve this by letting the probe atom feel the influence of the molecule, but not vice versa.
This allows us to define a local chemical potential $\Omega_{Z_\text{p}}(\mathbf{r})$ as the energy of the test charge of atom type $Z_\text{p}$ located at position $\rr$:
\begin{align*}
\x_p^{(t+1)} &= \mathbf{x}^{(t)}_i + \sum\limits_{j} \mathbf{v}^{(t)} \left(\mathbf{x}^{(t)}_j, \rr_{p}-\rr_{j} \right) \\
\Omega_{Z_\text{p}}(\mathbf{r}) &= f_\text{out}(\mathbf{x}^{(T)}_\text{p})
\end{align*}
It is important to note that this potential does not correspond to the actual potential of the molecule, but is a tool for us to visualize the spatial structure of the representation.

\begin{figure}[tb]
	\centering
	\includegraphics[width=\textwidth]{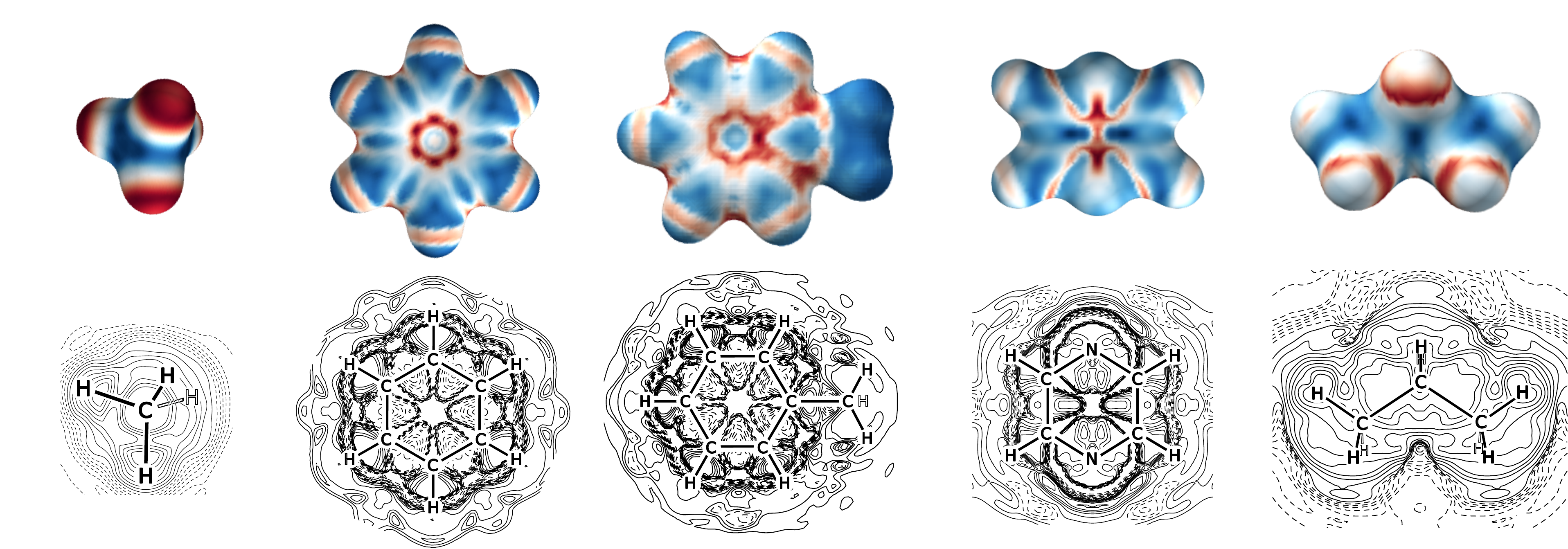}
	\caption{Local chemical potentials obtained with SchNet using a carbon probe for  methane, benzene, toluene, pyrazine and propane. They are shown on a $\sum_i \| \rr - \rr_i \|^{-2}=3.7${\AA}$^{-2}$ isosurface (top) as well as cuts through the center of the molecule (bottom). Dashed lines indicate regions of negative potential.\label{fig:lcp}}
\end{figure}

Fig.~\ref{fig:lcp} visualizes such local chemical potentials using a carbon probe for SchNet trained on QM9 on a smooth isosurface with constant $\sum_i \|\mathbf{r} - \mathbf{r}_i \|^{-2}$ around a selection of molecules from the dataset.
Furthermore, we show cuts through the local chemical potentials of the molecules as contour plots.

The potentials reflect the expected symmetries that stem from the rotational and translational invariance of SchNet.
The low- and high-energy regions on the iso-surfaces are clearly separated.
In the cuts, we observe a high sensitivity to the probe position (i.e. high density of contour lines) near the atom positions, which is most clearly visible for the molecules with aromatic rings.
Both of these findings indicate that the learned representation is localized, which coincides with chemical intuition.

Since our local chemical potentials inherit the locality of atom-wise explanations, they can be similarly used as a visually more intuitive alternative for attributing local relevance.
On top of that, the visualizations mirror chemical concepts such as bond saturation as well as different degrees of aromaticity.
This makes them a powerful analysis tool for the chemistry researcher.


While the local chemical potentials introduced above deliver valuable and chemically plausible visualizations of the learned representation, they can not correspond to the actual potential generated by the molecule.
This is because we are not able to introduce a real point charge for probing into the network, but have to resort to full atoms that would significantly disturb the molecule if we allowed it to influence the other atoms.
In contrast, we are able to use the latent partial charges learned during the prediction of dipole moments to obtain an approximation of the electrostatic potential (ESP) of the molecule.
The ESP offers insights into the spatial distribution of charges inside a molecule and indicates regions which are attractive or repulsive to the probe atom. This information can in turn be used to interpret e.g. reaction outcomes or coordination to other molecules. 

The ESP of a molecule is the potential energy experienced by a probe charge $q_0$ in the electric field of a molecule. Using the partial charges $\hat{q}$ obtained above, we can obtain a corresponding ESP
\begin{equation}
E(\mathbf{r}_0) = \sum^N_i \frac{\hat{q}_i q_0}{||\mathbf{r}_i-\mathbf{r}_0||_2},
\end{equation}
where $\mathbf{r}_0$ and $q_0$ are the position and charge of the probe and $\mathbf{r}_i$ and $\hat{q}_i$ are the positions and partial charge of atom $i$ of the molecule.
Here, the charge distribution of the molecule is approximated by atom-wise latent partial charges learned in order to predict the dipole moment.
Therefore, this approximation only models the part of the ESP that is relevant to describe the dipole of the molecule.

\begin{figure}[tbp]
	\centering
	\newcolumntype{S}{>{\centering\arraybackslash} m{.33\textwidth} }
	\begin{tabular}{cSSS}
		\rotatebox{90}{BP}&\includegraphics[width=0.22\textwidth]{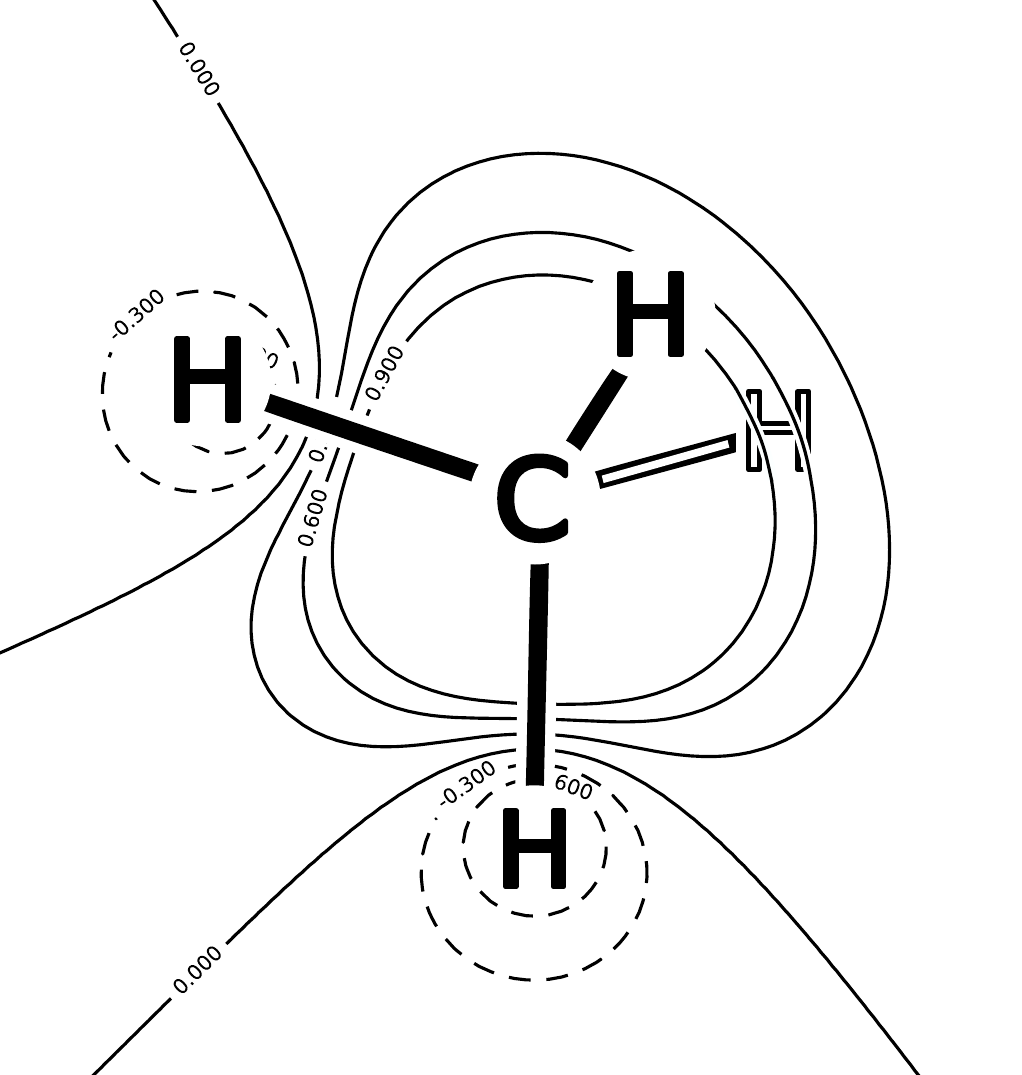} & \includegraphics[width=0.3\textwidth]{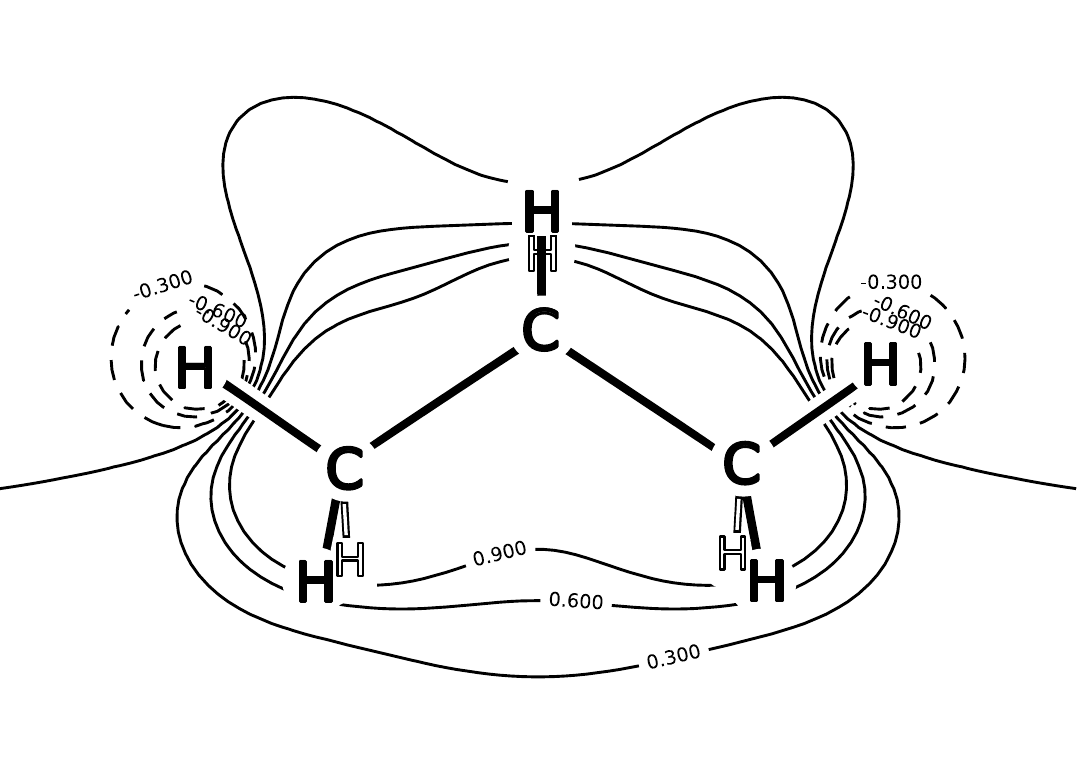} & \includegraphics[width=0.25\textwidth]{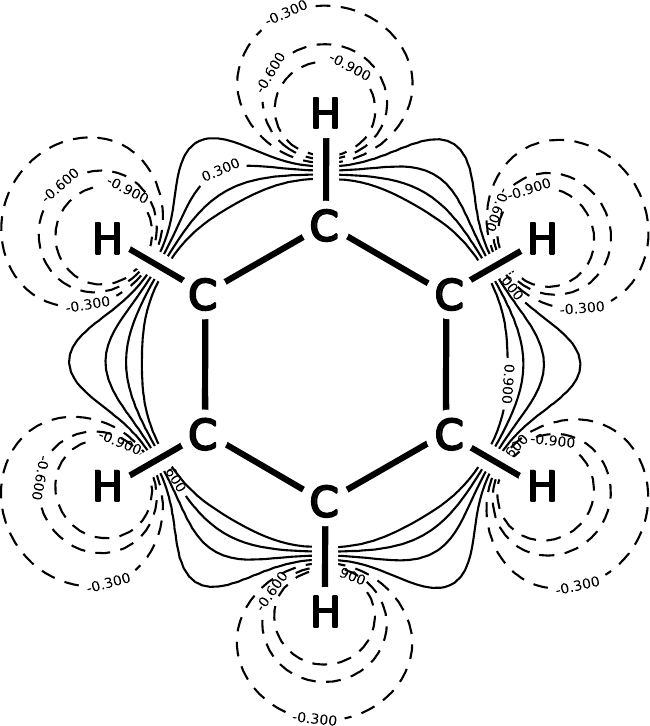} \\
		&Methane & Propane & Benzene \\
		\rotatebox{90}{SchNet}&\includegraphics[width=0.22\textwidth]{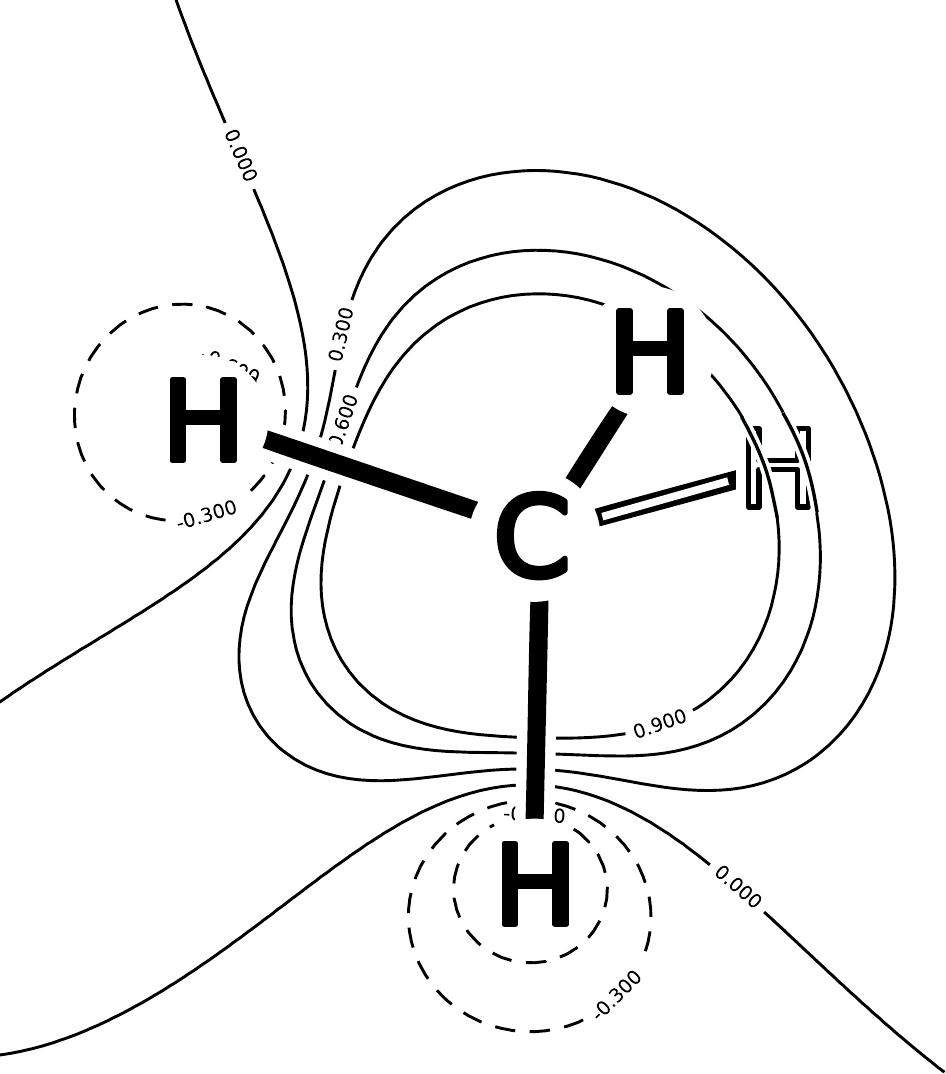} & \includegraphics[width=0.3\textwidth]{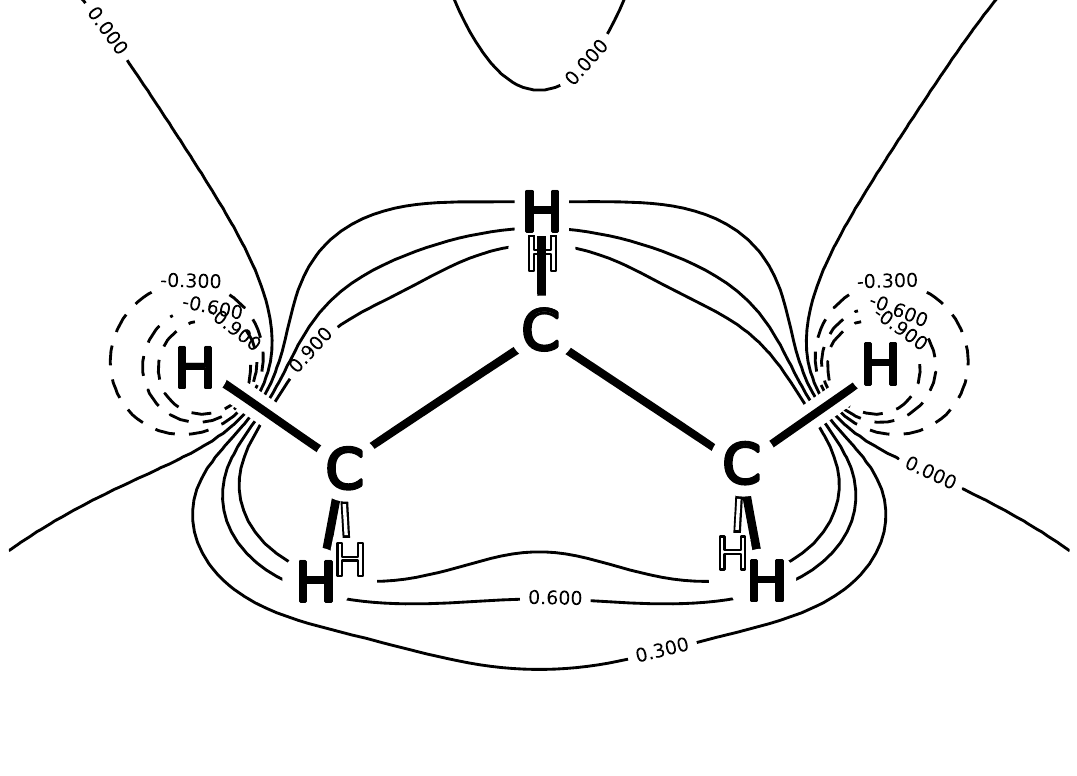} & \includegraphics[width=0.25\textwidth]{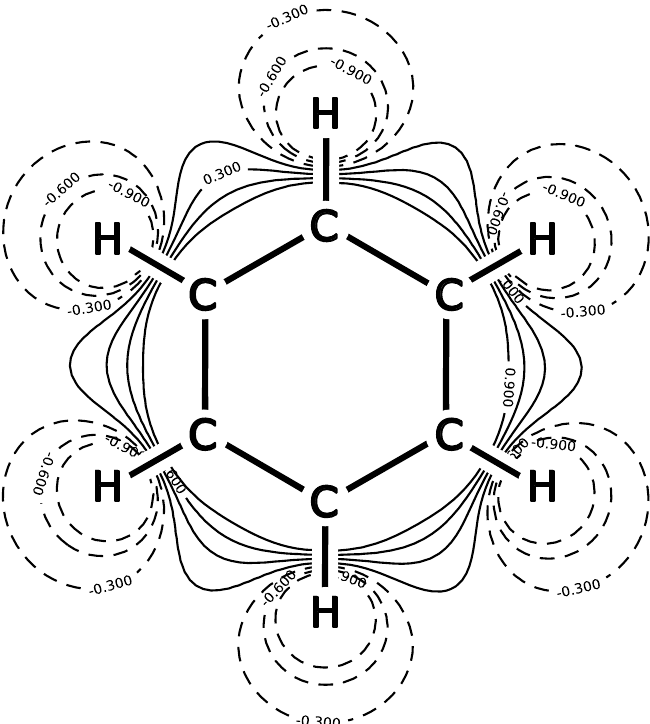} \\
		&&&\\
		\rotatebox{90}{BP}&\includegraphics[width=0.25\textwidth]{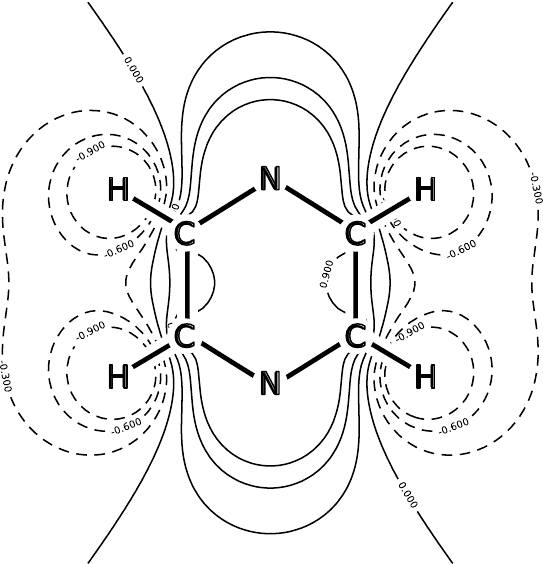} & \includegraphics[width=0.25\textwidth]{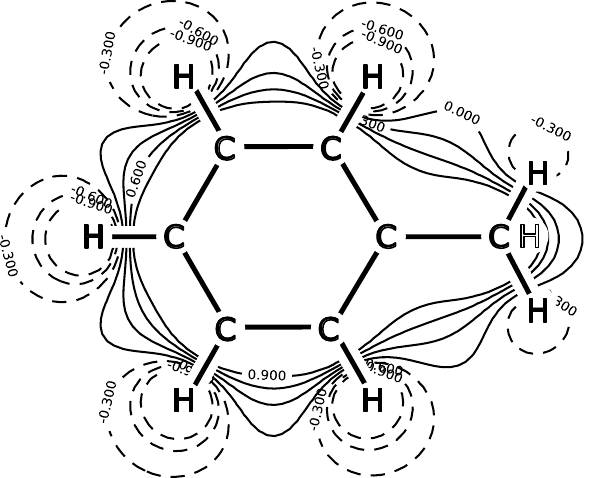} & \includegraphics[width=0.3\textwidth]{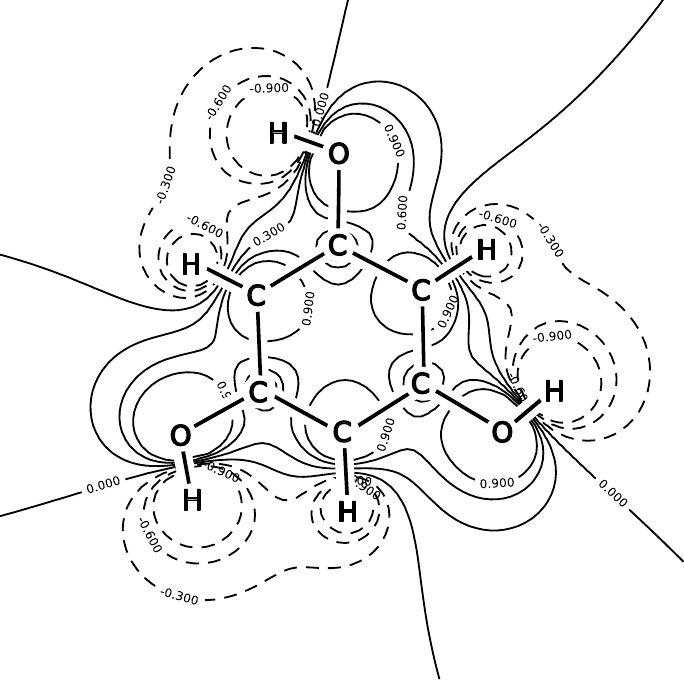} \\
		&Pyrazine & Toluene & Phloroglucinol \\
		\rotatebox{90}{SchNet}&\includegraphics[width=0.25\textwidth]{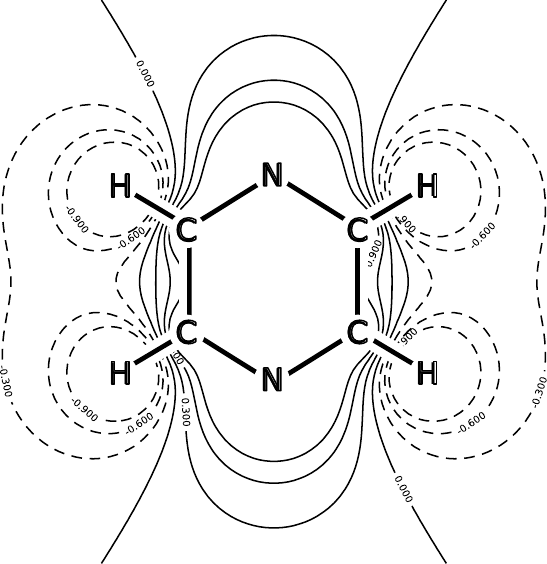} & \includegraphics[width=0.25\textwidth]{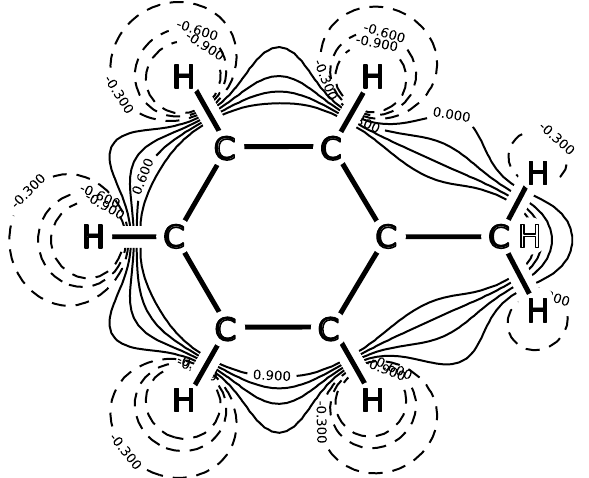} & \includegraphics[width=0.3\textwidth]{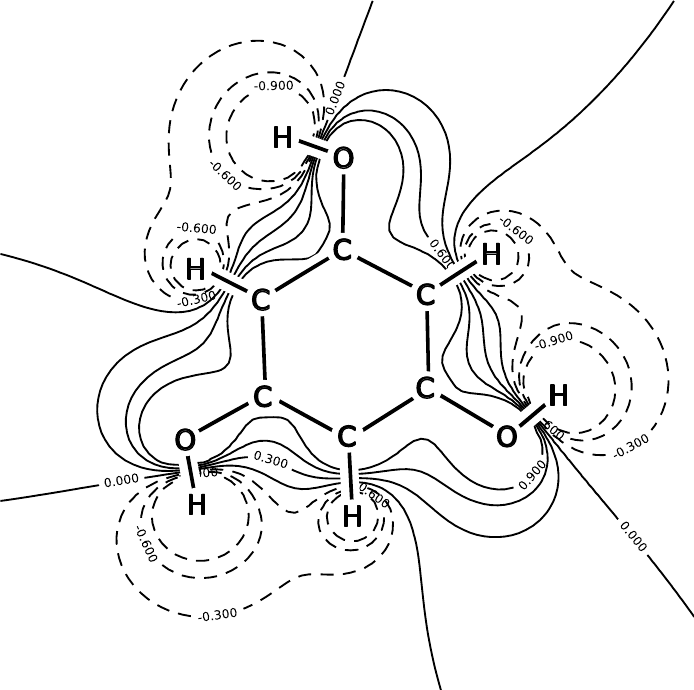} \\
	\end{tabular}
	\caption{Comparison of electrostatic potentials obtained with the atomic charges yielded by a BP type network and a SchNet and using a probe charge of $q_0=1$. Regions of positive potential are indicated with dashed lines. All charges are normalized.}
	\label{fig:esp}
\end{figure}
Figure~\ref{fig:esp} gives the ESPs of six molecules from QM9 as computed with latent partial charges from BP and SchNet.
Both models give very similar ESPs for the different molecules. 
This is a consequence of the similarity between the charge distributions produced by the different architectures (see Section~\ref{sec:distributions}) and further amplified by the damping introduced via the inverse dependence on the distance between probe and atoms. 
Looking at the ESPs in general, we find that the obtained maps show excellent agreement with basic chemical reasoning. 
In the molecules containing only hydrogen and carbon (methane, propane, benzene, toluene), one would expect the hydrogen atoms to carry a slight positive charge and hence lead to unfavorable interactions with the equally positively charged probe. 
The opposite holds true for the carbon atom.
This feature is indeed observed in all the ESP maps. 
In a similar manner, one would expect the oxygen atoms in phloroglucinol to carry a negative charge, due to their electron-withdrawing properties. 
Thus, the ESP should show a negative area around these atoms, which is indeed the case in the examined ESPs.

Similar to the local chemical potentials, the ESPs are a valuable tool for analyzing the obtained features.
Moreover, they are grounded in physics which makes them readily interpretable.
Hence, ESPs present a valuable tool for model validation and allow to directly extract spatially resolved chemical insights.


\subsection{Visualization of Chemically Resolved Insights}

\begin{figure}[tb]
	\centering
	\includegraphics[width=0.7\textwidth]{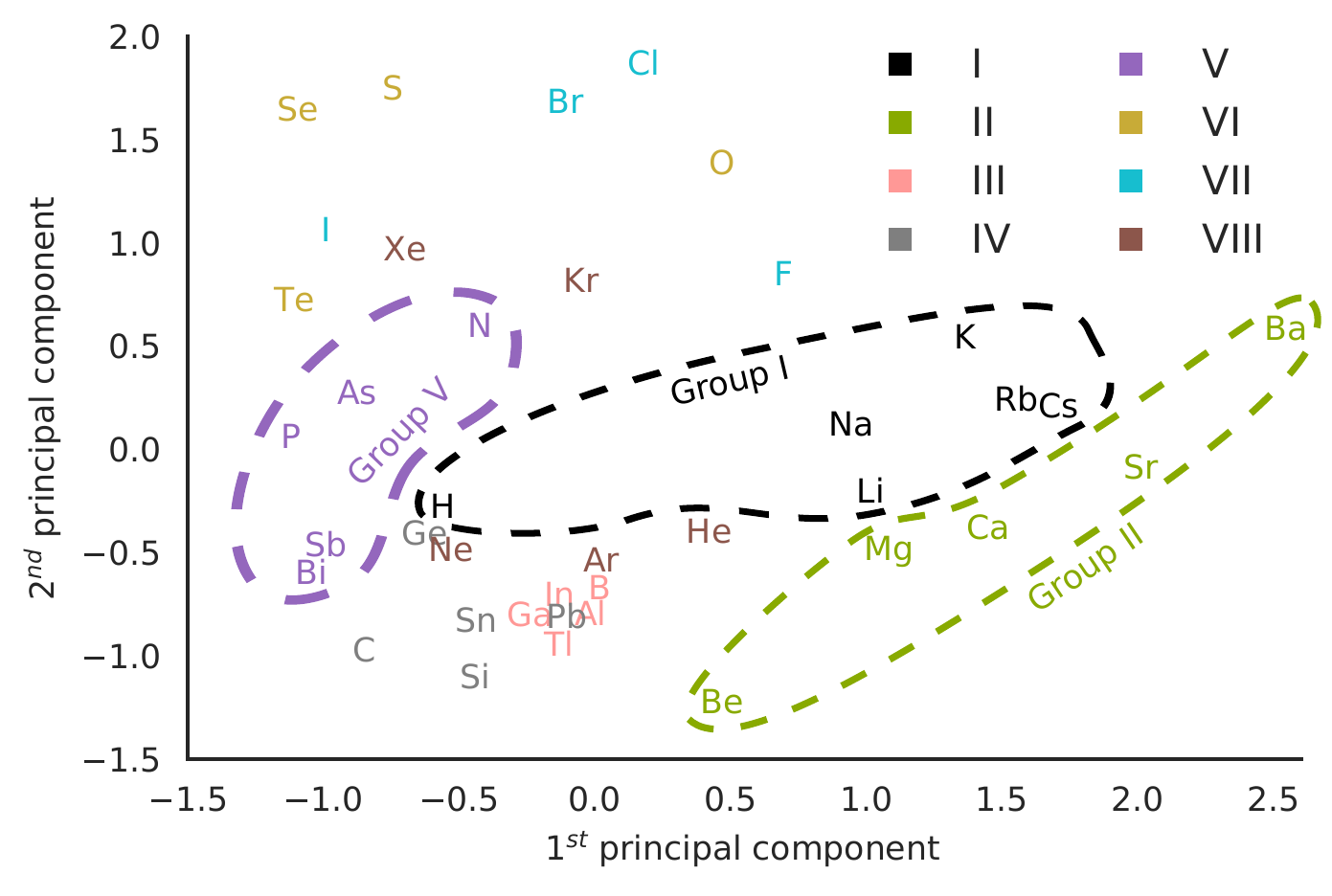}
	\caption{The two leading principal components of the learned embeddings $x_0$ of sp
		atoms learned by SchNet from the Materials Project dataset. We recognize a structure in the embedding space according to the groups of the periodic table (shown exemplary for groups I, II and V and color-coded online) as well as an ordering from lighter to heavier elements within the groups, e.g., in groups I and II from light atoms (left) to heavier atoms (right).\label{fig:embedding}}
\end{figure}

While a lot of handcrafted descriptors consider different atom types orthogonal~\cite{behler2007generalized,Hansen-JCPL,schutt2014represent} or use nuclear charges to encode atom similarities~\cite{rupp2012fast,gastegger2018wacsf}, SchNet and DTNN allows for cross-element generalization through the high-dimension embeddings of chemical elements~\cite{schutt2017quantum,schutt2018jcpschnet}
If the trained models learn to efficiently make use of this possibility, we should be able to extract element similarities from the embeddings that resemble chemical intuition.
Since QM9 only contains five atom types (H, C, N, O, F), we will perform this analysis on the Materials Project dataset of crystal structures as it includes 89 atom types ranging across the periodic table.

Fig.~\ref{fig:embedding} shows the two leading principal components of the element embeddings of the main group elements of the periodic table.
The projection explains only about 20\% of the variance, therefore atom types might appear closer than they are in the high-dimensional space.
However, we see that atoms belonging to the same group tend to form clusters.
This is especially apparent for main groups 1-5, while groups 6-8 appear to be slightly more scattered.
In group 1, hydrogen lies further apart from the other members which coincides with its special status, being the element without core electrons. 
Beyond that, there are partial orderings of elements according to their period within some of the groups.
There are orderings from light to heavier elements, e.g. in group 1 (left to right: H - [Na,Li] - [K, Rb, Cs]), group 2 (left to right: Be - Mg - Ca - Sr - Ba) and group 5 (top to bottom: N-[As, P]-[Sb,Bi]).

Note that these extracted chemical insights were not imposed by the SchNet architecture onto the embeddings as they were initialized randomly before training.
They had to be inferred by the model based on the co-occurrence of atoms in the crystal structures of the training data.

\section{Conclusions}

We have presented two atomistic neural networks that enable fast and accurate predictions of energies and dipole moments: Behler-Parrinello (BP) networks that use atom-centered symmetry functions as input features and the end-to-end architecture SchNet which learns representations of atomistic systems directly from first-principles.
In these architectures, chemical properties are modeled using physically motivated aggregation layers over atom-wise latent contributions.
At the same time, latent local contributions correspond to the assignment of atom-wise relevances in the spirit of LRP~\cite{Bach2015} or similar methods~\cite{Montavon2017,kindermans2018learning,Montavon2017}.
However, since the models are constrained to assemble the final target from atom-wise contributions in the forward pass, we do not have to resort to relevance redistribution techniques.
On this ground, we have presented various interpretation techniques to extract insights about the learned representations as well as the underlying quantum-chemical problems.

Both examined models are able to obtain consistent partitionings of the energy -- a major challenge for quantum-mechanical calculations.
Particularly remarkable is the possibility to obtain chemically plausible rankings of aromatic rings regarding their stability.
Using a virtual probe atom, we are able to extend atom-wise energy contributions to visualizations in 3-d space in the form of local chemical potentials.
These further improve interpretability of the energy partitioning and resemble chemical concepts such as bond saturation, electronegativity and different degrees of aromaticity.
In the same spirit, we have examined latent partial charges obtained during the prediction of dipole molecular moments.
They allow us to visualize the approximate charge distribution of the molecule using electrostatic potentials, which are grounded in physics and show excellent agreement with basic chemical intuition. 
Both local chemical potentials as well as electrostatic potentials present a valuable tool for model validation as well as extracting spatially resolved chemical insights.
Finally, we have examined embeddings of chemical elements obtained from training SchNet on a diverse set of crystal structures.
The obtained embeddings recover knowledge about chemical elements present in the structure of the periodic table.
This guides the way to future work, extending the analysis to measure chemical similarity of local structures.

While accurate predictions are a necessary requirement for every machine learning model in quantum chemistry, it is crucial that the model is able to facilitate new research.
Here, interpretability constitutes an essential building block for researcher in the respective field to validate, understand and ultimatively trust the machine learning model.
Therefore, interpretation techniques should be closely oriented towards analysis methods familiar to the respective field, lowering the initial barrier for researchers unfamiliar with non-linear models of machine learning.
For the same reason, it is beneficial if the interpretable properties are directly obtained during the forward pass.
This ensures that they are ground truth -- i.e. they are assembled to form the final prediction using a known functional form -- without having to rely on an approximate redistribution of relevances.
The excellent agreement of the examined representations with chemical knowledge is a clear demonstration of the ability of atomistic neural networks to open up new venues for data-driven research in the chemistry, physics and materials science.

\section*{Acknowledgements}
This work was supported by the Federal Ministry of Education and Research (BMBF) for the Berlin Big Data
Center BBDC (01IS14013A). Additional support was provided by the DFG (MU 987/20-1), from the European
Union’s Horizon 2020 research and innovation program under the Marie Sklodowska-Curie grant agreement NO 792572, the BK21 program funded by Korean National Research Foundation grant (No. 2012-005741) and the Institute for Information \& Communications Technology Promotion (IITP) grant funded by the Korea government (no. 2017-0-00451). A.T. acknowledges support from the European Research Council (ERC-CoG grant BeStMo).

\bibliographystyle{splncs03_unsrt}
\bibliography{references}

\end{document}